\documentclass[psfig,pstricks,twocolumn,showpacs,preprintnumbers,amsmath,amssymb,superscriptaddress]{revtex4-1}

\usepackage{graphicx}          
\usepackage{dcolumn}           

\usepackage{physics}

\usepackage{array,booktabs,lmodern}
\usepackage[sc]{mathpazo}
\usepackage[scaled=1.00]{helvet}
\usepackage[final]{listings,microtype}
\usepackage[T1]{fontenc}
\usepackage[version=3]{mhchem} 
\usepackage{siunitx} 
\usepackage[english]{babel}
\usepackage{float}
\usepackage{natbib}
\usepackage{setspace}
\usepackage{xkeyval}
\usepackage{natmove}
\usepackage{titlesec}
\newcommand{\mat}[1]{\mbox{\boldmath{$#1$}}} 
\usepackage{xfrac}
\usepackage{relsize}
\usepackage{hyperref}
\hypersetup{
    colorlinks=true,
    urlcolor= blue,
    citecolor=blue,
    linkcolor= blue,
    bookmarksopen=false,
    }

\usepackage[title]{appendix}

\begin{document}

\preprint{Page}

\title{Boosting the efficiency of {\it{ab initio}} electron-phonon coupling calculations through dual interpolation} 

\author{Anderson S. Chaves}
\affiliation{John A. Paulson School of Engineering and Applied Sciences, Harvard University, Cambridge, Massachusetts, 02138, USA}
\affiliation{Gleb Wataghin Institute of Physics and Center for Computing in 
Engineering \& Sciences, University of Campinas, PO Box 13083-859, Campinas, SP, Brazil}
\author{Alex Antonelli}
\affiliation{Gleb Wataghin Institute of Physics and Center for Computing in 
Engineering \& Sciences, University of Campinas, PO Box 13083-859, Campinas, SP, Brazil}
\author{Daniel T. Larson}
\affiliation{Department of Physics, Harvard University, Cambridge, Massachusetts, 02138, USA}
\author{Efthimios Kaxiras}
\affiliation{Department of Physics, Harvard University, Cambridge, Massachusetts, 02138, USA}
\affiliation{John A. Paulson School of Engineering and Applied Sciences, Harvard University, Cambridge, Massachusetts, 02138, USA}

\date{\today}

\begin{abstract}

The coupling between electrons and phonons in solids
plays a central role in describing many phenomena, including 
superconductivity and thermoelecric transport. Calculations of this 
coupling are exceedingly demanding as they necessitate 
integrations over both the electron and phonon momenta, 
both of which span the Brillouin zone of the crystal, independently. 
We present here an {\it{ab initio}} method for efficiently 
calculating electron-phonon mediated transport properties 
by dramatically accelerating the computation of the double integrals 
with a dual interpolation technique that combines 
maximally localized Wannier functions with symmetry-adapted plane 
waves. The performance gain in relation to the current 
state-of-the-art Wannier-Fourier interpolation is approximately $2n_s \times M$, 
where $n_s$ is the number of crystal symmetry operations and $M$,
a number in the range $5-60$, governs the expansion in star functions. 
We demonstrate with several examples how our method 
performs some {\it{ab initio}} calculations involving electron-phonon interactions.
\end{abstract}


\maketitle

The electron-boson coupling is ubiquitous 
in physical phenomena through the whole spectrum of the physics of solids.
In particular, electron-phonon ({\it{el-ph}})
coupling plays a fundamental role in the renormalization of electronic
and vibrational energy scales, thus determining the coupling itself,
with important consequences for transport 
properties\cite{giustino2017electron,marini2008ab,park2007velocity,restrepo2009first}.
Conventional superconductivity
is a case in point, where the interactions between electrons 
and phonons give rise to Cooper pairing.\cite{margine2014two}
Other examples include the temperature dependence of electronic conductivity
and thermoelectric transport properties,\cite{fiorentini2016thermoelectric,wang2011thermoelectric} 
as well as phonon-assisted optical absorption
in indirect-gap semiconductors.\cite{noffsinger2012} 
Interest in thermoelectrics has increased rapidly in recent years, 
partly due to the expectation of discovering higher figure-of-merit materials,
boosted by the nanotechnology revolution. \cite{boukai2011silicon}
A major goal of theory has been to predict thermoelectric transport properties 
directly from atomistic-scale calculations without any adjustable or 
empirical parameters,\cite{restrepo2009first,wang2011thermoelectric,fiorentini2016thermoelectric} 
particularly combining density functional theory (DFT) and many-body perturbation theory. 
Despite great advances, such calculations remain very demanding 
and still pose a challenge, even for simple crystalline bulk systems.

\begin{figure*}
\centering
\includegraphics[angle=270,origin=c,width=1\textwidth]{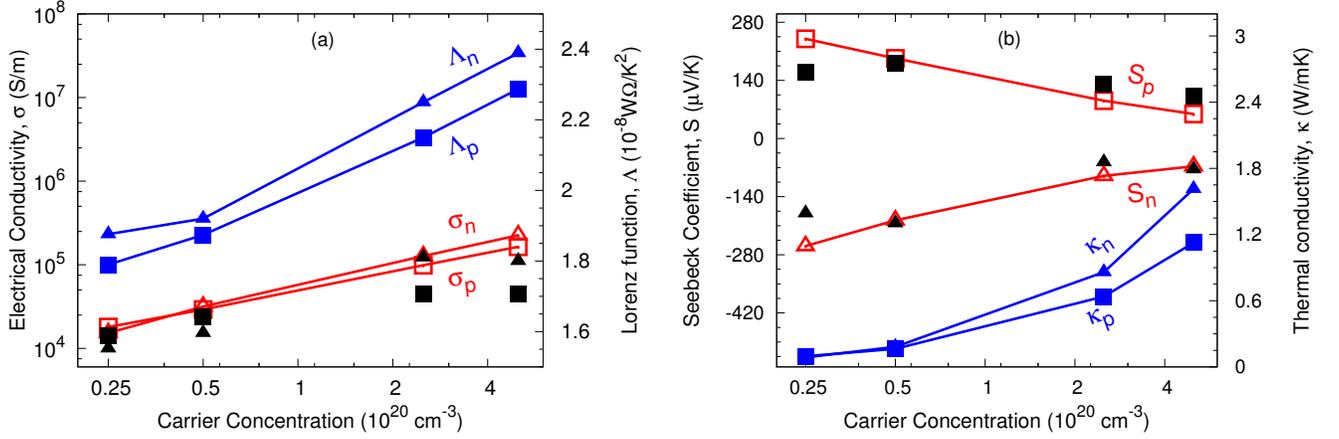}
\vspace*{-62mm}\caption{\label{fig1}
Thermoelectric transport properties ${\it{TP}}$ for p- and n-type 
doped Si polycrystals: (a) Electrical
conductivity, $\sigma$, (red) compared with experimental values\cite{strasser2004micromachined} (filled black symbols)
and Lorenz function, $\Lambda$, (blue); (b) Seebeck coefficient, S, 
(red) compared with experimental data\cite{strasser2004micromachined} 
(filled black symbols) and
thermal conductivity due to the carriers, $\kappa$, (blue) calculated from the relaxation times 
due to scatterings by {\it{el-ph}} coupling and ionized impurities (see text for details).} 
\end{figure*}
A well-established approach, using 
a first-principles description of {\it{el-ph}} coupling, relies on
solving the {\it{el-ph}} matrix elements through density-functional perturbation theory
(DFPT)\cite{baroni2001phonons}.
The {\it{el-ph}} matrix elements, 
$g({\bf{k}},{\bf{k+q}}) =(\bra{{\bf{k}}+{\bf{q}}} {\delta}_{{\bf{q}},\beta} V^{KS} \ket{{\bf{k}}})_{uc}$, 
correspond to the electronic scattering 
calculated from the variations of the Kohn-Sham (KS) potential 
due to phonon perturbations with wavevector $\bf{q}$
and branch index $\beta$ within the unit cell ($uc$).
To obtain transport properties ({\it{TP}}), 
$|g(\bf{k},\bf{k+q})|^2$ must
be integrated over the electron and phonon momenta,
both spanning the entire Brillouin Zone (BZ), independently. 
This double integration requires very
fine sampling of the electron and phonon wavevectors to achieve
numerical convergence, 
which represents the bulk of the computational burden.  
The application of crystal symmetry properties for the full integration
of {\it{TP}}, which depend directly on the {\it{el-ph}} matrix elements, is not allowed. 
Even if the wavevector ${\bf{k}}$ in the {\it{el-ph}} matrix element lies
within the symmetry-reduced portion  (irreducible wedge) of the BZ, 
the transferred momenta ${\bf{k + q}}$ spread out in the whole zone
because ${\bf{q}}$ belongs to an uniform mesh.
Dense sampling of the BZ is prohibitive, the reason being 
the connection between transferred momenta 
with equally dense ${\bf{k}}$-point meshes.

Specialized numerical
techniques have been developed to address this problem.
One attempt to simplify the brute-force integration is based on
pre-screening of subsets of the reciprocal space, such as relevant conduction
pockets within the neighborhood of band extrema that significantly contribute to
the integral.
Alternatively, interpolation schemes, such as
linear\cite{li2015electrical} or 
Wannier-based\cite{giustino2007electron,giustino2017electron} ones, have been developed
to improve convergence. 
In particular, the interpolation of the {\it{el-ph}} matrix elements on the
basis of Wannier functions introduced by
Giustino, Cohen, and Louie,\cite{giustino2007electron} has proven  
very successful in calculating
properties with more favorable scaling than using directly the DFPT approach.
In this Letter, we present a novel method for the computation of {\it{el-ph}} 
mediated {\it{TP}} which uses two interpolations: the first one is the 
usual Wannier-Fourier (W-F) interpolation, followed by a 
second one based on symmetry-adapted
plane-waves (PW). Our method leads to an efficient sampling of 
extremely fine, homogeneous ${\bf{k}}$ and ${\bf{q}}$ grids, with a 
significant decrease in computational cost compared to W-F calculations 
with a single interpolation.   

\begin{figure}
\centering
\includegraphics[angle=270,origin=c,width=0.5\textwidth]{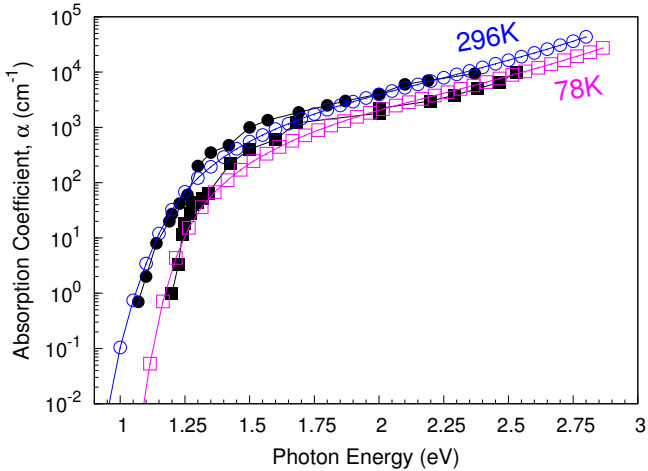}
\vspace*{-16mm}\caption{\label{fig2}
Phonon-assisted optical absorption, $\alpha$, of Si at $296$K and $78$K 
as a function of photon energies. 
Our results (blue and magenta) are in good agreement 
with corresponding experimental results (black)\cite{braunstein1958intrinsic}.}
\end{figure}

To illustrate the capability of our method,
we considered realistic properties of solids (see SM for more details).
Fig.~\ref{fig1} shows the
calculated thermoelectric $\it{TP}$ for Si polycrystals using
our dual interpolation method to calculate the relaxation time within 
the Boltzmann transport equation (BTE).
We calculated the electrical
conductivity, $\sigma$, Seebeck coefficient, S, Lorenz function, $\Lambda$,
and thermal conductivity due to the carriers, $\kappa$, as functions of carrier
concentration. Our results for $n$ and $p$-doped Si polycrystals agree reasonably well
with available experimental $\it{TP}$. For these calculations we also included
the scattering by ionized impurities, within the Brooks-Herring theory (see SM), considering
in all calculations a fixed unitary ratio between impurities and carrier concentrations, which
are the only input parameter along with the crystal structure.
Fig.~\ref{fig2} shows the results of phonon-assisted optical absorption for Si
at $296$K and $78$K as a function of the photon energies. Our calculations, based on
our dual interpolation method and the theory developed by Hall, Bardeen and
Blatt (see \cite{noffsinger2012}), are in good agreement
with experimental results. 

Before presenting our method we review the basic concept and 
analyze the advantages and drawbacks of W-F interpolation.
Using W-F interpolation, the {\it{el-ph}} matrix elements 
$g(\mathbf{k},\mathbf{q})$ can be calculated on coarse $\mathbf{k},\mathbf{q}$ 
meshes and then interpolated onto much finer $\mathbf{k}',\mathbf{q}'$ 
meshes through simple matrix multiplication.\cite{giustino2007electron} 
The matrix elements on the fine mesh are given by
\begin{equation} 
g({\bf{k^ {\prime}}},{\bf{q^ {\prime}}}) = \frac{1}{N_e} \sum_{{\bf{R}}_e,{\bf{R}}_p} e^{i({\bf{k^{\prime}}}\cdot{\bf{R}}_e+{\bf{q^{\prime}}}\cdot{\bf{R}}_p)}{\bf{U}}_{{\bf{k^{\prime}}}+{\bf{q^{\prime}}}}g({\bf{R}}_e,{\bf{R}}_p){\bf{U}}_{{\bf{k^{\prime}}}}^{\dagger} {\bf{u}}_{\bf{q^{\prime}}},
\label{g_interpolation}
\end{equation}
where ${\bf{R}}_{e}$ and ${\bf{R}}_{p}$ are primitive lattice vectors 
of the Wigner-Seitz (WS) supercell with Born-von-K{\'a}rm{\'a}n (BvK) periodic boundary conditions,  
{\bf{U}}$_{{\bf{k^{\prime}}}}$ ({\bf{u}}$_{{\bf{q^{\prime}}}}$) is a diagonalizer matrix
over ${\bf{k^{\prime}}}$ (${\bf{q^{\prime}}}$) indices from Wannier to Bloch representations
for electrons (phonons) 
and the {\it{el-ph}} matrix elements in the Wannier representation are given by
\begin{equation}
g({\bf{R}}_e,{\bf{R}}_p) = \frac{1}{N_p} \sum_{{\bf{k}},{\bf{q}}} e^{-i({\bf{k}}\cdot{\bf{R}}_e+{\bf{q}}\cdot{\bf{R}}_p)}{\bf{U}}_{{\bf{k}}+{\bf{q}}}^{\dagger}g({\bf{k}},{\bf{q}}){\bf{U}}_{{\bf{k}}} {\bf{u}}_{{\bf{q}}}^ {-1}~,
\end{equation}
{\bf{U}}$_{{\bf{k}}}$ is a unitary matrix corresponding to
the rotation of the corresponding electronic states from
Bloch to Wannier representations within the gauge of
maximally localized Wannier functions (MLWF),\cite{marzari1997maximally}
and {\bf{u}}$_{{\bf{q}}}$ is a unitary rotation matrix
from Bloch to MLWF for phonons.
The strength of the W-F interpolation method is the fact that
one only needs to perform calculations over the initial
coarse $\bf{k},\bf{q}$ meshes, and then can use 
Eq.~(\ref{g_interpolation}) to determine $g(\bf{k}',\bf{q}')$ on 
finer $\bf{k^{\prime}},\bf{q^{\prime}}$ meshes. 
For this,  we neglect 
the matrix elements outside
the WS supercell generated from the initial coarse BZ mesh.

The accuracy of W-F calculations strongly depends
on the spatial localization of $g({\bf{R}}_e,{\bf{R}}_p)$ within Eq.~(\ref{g_interpolation}). 
A more detailed analysis suggests $g({\bf{R}}_e,{\bf{R}}_p)$
should decay in the variable ${\bf{R}}_e$ at least with the rapidity of MLWFs.
For ${\bf{R}}_e = 0$, $g(0,{\bf{R}}_p)$ decays 
with ${\bf{R}}_p$ due to the screened Coulomb interaction 
of the dipole potential generated by 
atomic displacement. Thus the localization of $g({\bf{R}}_e,{\bf{R}}_p)$ 
depends strongly on the dielectric properties of the system. 
In particular, Friedel oscilations\cite{fetter2012quantum} ($|{\bf{R}}_p|^{-3}$) 
and quadrupole behavior\cite{pick1970microscopic} 
($|{\bf{R}}_p|^{-4}$) are intimately related to the screening properties 
of metals and nonpolar semiconductors, respectively.

Despite the advantages of the W-F interpolation
and its more favorable scaling, there are still some drawbacks.
The method is computationally intensive when 
many $\bf{k}$/$\bf{q}$ points are needed to achieve converged values for {\it{TP}}.
The main computational operations are the simple matrix 
multiplications shown in Eq.~\ref{g_interpolation}, with a
computational complexity of $\mathcal{O}(n^3)$ for classical
computation, where $n$ is the matrix size involved. For final 
dense grids with $N_{f}^{{\bf{k^{\prime}}}}$ ($N_{f}^{{\bf{q^{\prime}}}}$) 
${\bf{k^{\prime}}}$ (${\bf{q^{\prime}}}$) points, 
the number of floating-point operations 
reach $\approx N_{f}^{\bf{k^\prime}} \times N_{f}^{\bf{q^\prime}} n^3$.
To reduce the computational
cost of Wannier-based calculations, some strategies 
have been adopted, including double grid schemes\cite{fiorentini2016thermoelectric}
or (quasi) Monte-Carlo (MC) integrations.\cite{ponce2014verification} 
In the former, only the bandstructure and phonon dispersion
are calculated over an ultrafine grid ($N_f \times N_f \times N_f$), 
while {\it{el-ph}} matrix elements are computed over
a moderate grid ($N_{el-ph} \times N_{el-ph} \times N_{el-ph}$, 
with $N_{el-ph} = s \times N_f$ and $s = 1/2, 1/3$)
and extrapolated to the ultrafine grid 
assuming that the {\it{el-ph}} coupling function is smooth.
The drawbacks here consist of the extrapolation which is often fraught with risk,
and the very modest gain factor of the method.
On the other hand, by using (quasi) MC integration, 
one has to test very dense sets of random (or quasi-random) 
$\bf{k}$ or $\bf{q}$-points, which is a serious drawback.

Our method proceeds as follows. For clarity, we describe 
the procedure for doing the partial $\mathbf{q}'$ integration first, 
but one can easily switch the order of integration. We begin by 
computing $g(\mathbf{k},\mathbf{q})$ over coarse $\mathbf{k}$ and $\mathbf{q}$ 
meshes. Next, using W-F interpolation, we determine $g$ over a 
finer $\mathbf{q}'$ mesh and perform the partial integration at 
each of the $n_\mathbf{\bar{k}}$ irreducible $\mathbf{k}$-points, ${\bf{\bar{k}}}_l$, 
corresponding to a moderate regular $\mathbf{k}$-mesh (${\bf{k^r}}$).
Thus we obtain a function
$f({\bf{\bar{k}}}_l) \propto \int_{BZ} |g({\bf{\bar{k}}}_l + {\bf{q^ {\prime}}},{\bf{\bar{k}}}_l)|^2 d{{\bf{q^ {\prime}}}}$
containing first-principles {\it{el-ph}} coupling properties defined 
at selected high-symmetry points in the corresponding $\mathbf{k}$-space.
Given $f(\mathbf{\bar{k}}_l)$, the next goal is to find a smooth 
interpolation over the whole $\mathbf{k}$-space on a finer grid. 
Such an interpolation problem may present severe difficulties
if one uses an inadequate basis set. We show below that 
such an interpolation over the entire BZ,
combined with periodic or other boundary conditions, can be properly
constructed from a basis set possessing the appropriate
crystal symmetry. 

As the full symmetry of the crystal's reciprocal space
is contained in the function $f({\bf{\bar{k}}}_l)$,
it is natural to use symmetry-adapted PW or star functions, 
$\Upsilon_m({\bf{k^{\prime}}})$,
as a basis set to Fourier expand $f$~\cite{chadi1973special}:
\begin{equation} \label{expansion}
\tilde{f}({\bf{k^ {\prime}}}) = \sum_{m=1}^M a_m \Upsilon_m({\bf{k^{\prime}}})~,
\end{equation}
where
$\Upsilon_m({\bf{k^ {\prime}}})=\frac{1}{n_s}\sum_{\{\upsilon\}}\exp({i(\upsilon {\bf{R}}_m)\cdot{\bf{k^{\prime}}}})~,$
with the sum running over all $n_s$ point group symmetry operations $\{\upsilon\}$
on the direct lattice translations, ${\bf{R}}_m$.
Star functions obey orthogonality relations involving BZ summations\cite{chadi1973special}, 
are totally symmetric under all point-group operations, and are ordered 
such that the magnitude of $\bf{R_m}$ is nondecreasing as $m$ increases, 
defining each star function to a given shell of lattice vectors.
By taking into account the symmetry, it is expected that
this expansion would converge much faster than using a regular Fourier expansion.
Following the approach proposed by Shankland-Koelling-Wood\cite{Shankland,koelling1986interpolation}, 
we take the number of star functions in the expansion, $M$,
to be greater than the number of data points ($M > n_{{\bf{\bar{k}}}}$). 
We then require the fit function, $\tilde{f}$, to pass through 
the data points and use the extra freedom from 
additional basis functions to minimize a 
spline-like roughness functional in order to suppress oscillations 
between data points, resulting in a well behaved function throughout the BZ.  

We adopt the spline-like roughness functional
defined by Pickett, Krakauer and Allen~\cite{pickett}, 
\begin{equation}
\label{R}
\Pi = \sum_{m=2}^M \lvert a_m \lvert ^2 \rho(R_m)
\end{equation}
with
$\rho(R_m) = \left(1-c_1\left({\frac{R_m}{R_{min}}}\right)^2\right)^2+c_2(\frac{R_m}{R_{min}})^6~,$
where $R_m = \lvert {\bf{R}}_m \lvert$, $R_{min}$ is the magnitude
of the smallest nonzero lattice vector, and $c_1 = c_2 = 3/4$.
Such a functional is more physically appealing
than the original functional proposed by Shankland-Koelling-Wood, in the sense that
departures of $\tilde{f}$ is minimized from its mean value, $a_1$, instead of zero.
The main problem in the expansion by star functions in Eq.~(\ref{expansion})
is the determination of the Fourier coefficients, $a_m$.
Thus, a Lagrange multiplier method can be used toward this goal, 
once the problem has been reduced to
minimizing ${\Pi}$ subject to the constraints,
$\tilde{f}({\bf{\bar{k}}}_l)={f}({\bf{\bar{k}}}_l)$,
in relation to $a_m$.
Consequently, the result of this minimization is
\begin{equation}
a_m = \begin{cases}
\rho(R_m)^{-1} \sum_{l=1}^{n_{\bf{\bar{k}}}-1} \lambda^*_l\left[\Upsilon_m^{*}({\bf{\bar{k}}}_l) - \Upsilon_m^{*}({\bf{\bar{k}}}_{n_{{\bf{\bar{k}}}}})\right],  &  m>1, \\
f({\bf{\bar{k}}}_{n_{\bf{\bar{k}}}})-\sum_{m=2}^M a_m \Upsilon_m ({\bf{\bar{k}}}_{n_{{\bf{\bar{k}}}}}),  &  m=1,
\end{cases}
\end{equation}
in which the Lagrange multipliers, $\lambda^*_l$, can be evaluated from
\begin{equation}
f({\bf{\bar{k}}}_p) - f({\bf{\bar{k}}}_{n_{{\bf{\bar{k}}}}}) = \sum_{l=1}^{n_{{\bf{\bar{k}}}}-1}{\bf{H}}_{pl}\lambda^*_l~,
\end{equation}
with
\begin{equation}
{\bf{H}}_{pl} = \sum_{m=2}^M \frac{\left[\Upsilon_m({\bf{\bar{k}}}_p) - \Upsilon_m({\bf{\bar{k}}}_{n_{{\bf{\bar{k}}}}})\right]\left[\Upsilon_m^{*}({\bf{\bar{k}}}_l)-\Upsilon_m^{*}({\bf{\bar{k}}}_{n_{{\bf{\bar{k}}}}})\right]}{\rho(R_m)}~,
\end{equation}
a positive-definite symmetric matrix that can be determined once 
for a given crystal problem and can be easily crafted numerically.

\begin{figure*}
\centering
\includegraphics[width=0.98\textwidth]{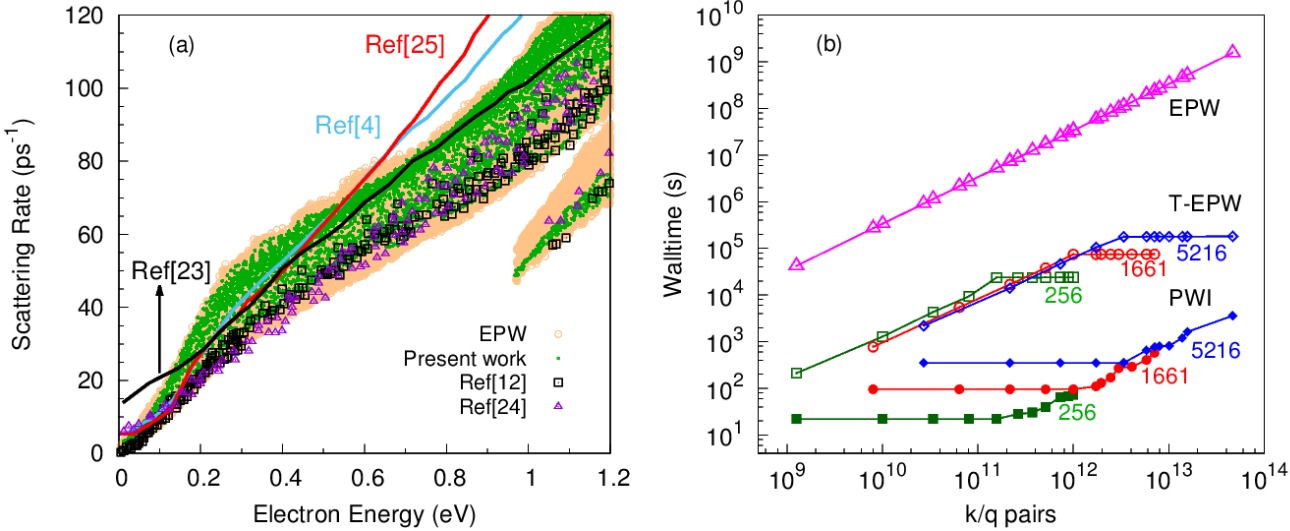}
\vspace*{6mm}\caption{\label{fig3}
(a) Scattering rates for Si at $300$ K calculated with the dual interpolation
method (green dots), DFT with linear interpolation
(black\cite{sun2012intrinsic} and light blue\cite{restrepo2009first} lines and black
squares\cite{li2015electrical}), W-F interpolation using EPW
(purple triangles\cite{qiu2015first}
and light orange circles),
and tight-binding calculations (red\cite{Rideau} line).
The W-F calculations
used $(30)^3$ {\bf{k}}/$(60)^3$ {\bf{q}}
($(100)^3$ {\bf{k}}/$(40)^3$ {\bf{q}}) meshes in
calculations represented by the purple triangles (orange circles).
(b) Walltime required to perform calculations for the electron self-energy due
to {\it{el-ph}} coupling in the Fan-Migdal approximation, for different grid sizes.
Empty magenta triangles correspond to the direct
calculation using EPW over homogeneous grids.
Empty (filled) green squares, empty (filled) red circles
and empty (filled) blue diamonds correspond
to the calculations using T-EPW (only second PW interpolation), starting from
$256$, $1661$ and $5216$ irreducible {$\mathbf{k}$} points, respectively
(see text for details).}
\end{figure*}

Once the Fourier coefficients are determined, a representation of $\tilde{f}$ is generated,
which can be written more clearly as a linear mapping of the W-F data,
\begin{equation}
\tilde{f}({\bf{k^{\prime}}}) = \sum_{l=1}^{n_{{\bf{\bar{k}}}}-1}J({\bf{\bar{k}}}_l,{\bf{k^{\prime}}}) [f({\bf{\bar{k}}}_l) - f({\bf{\bar{k}}}_{n_{{\bf{\bar{k}}}}})]~,
\end{equation}
where $J$ is the interpolation formula given by 
\begin{equation}
J({\bf{\bar{k}}}_l,{\bf{k^{\prime}}}) = \sum_{p=1}^{n_{{\bf{\bar{k}}}}-1}\sum_{m}^{M}\frac{[\Upsilon_{m}^{*}({\bf{\bar{k}}}_p)-\Upsilon_{m}^{*}({\bf{\bar{k}}}_{n_{{\bf{\bar{k}}}}})]\Upsilon_{m}({\bf{k^{\prime}}})}{\rho(R_m){\bf{H}}_{pl}}~,
\end{equation}
which transforms one ${\bf{k}}$-mesh into another one, that is, 
${\bf{\bar{k}}}_l \rightarrow {\bf{k^{\prime}}}$. This is the main result of our approach,
which allows great computational savings by transforming the W-F data
obtained over the ${\bf{k}}$-mesh of irreducible points (${\bf{\bar{k}}}_l$) 
into a homogeneous dense grid (${\bf{k^{\prime}}}$) that is larger than
the regular grid (${\bf{k^{r}}}$) that generates such irreducible points.
One important point to stress is that $J$ does not depend on data, but
it is completely defined by the lattice, namely the set of irreducible 
sampling (${\bf{\bar{k}}}_l$), the number of star functions ($M$), and
the form of spline-like roughness functional ($\Pi$).
In practice, in order to get a denser mesh, we rely on a Fast Fourier Transform (FFT) 
from the real space to the reciprocal space in order to compute  
the expansion given in Eq.~(\ref{expansion}). 
We take advantage of the BvK periodic boundary conditions to 
increase the real space by the expansion factor, $M$, as will be explained below, 
to get proportionally a new homogeneous ${\bf{k^{\prime}}}$-mesh finer than the original one.
As a result we get the full integration over 
very fine ${\bf{k^ {\prime}}}$ and ${\bf{q^ {\prime}}}$ meshes 
in order to calculate transport properties, that is, 
$TP \propto \sum_{{\bf{k^{\prime}}}} \tilde{f}({\bf{k^ {\prime}}})~.$

Our implementation for the second interpolation is based on modifications 
and adaptations of some subroutines of the BoltzTraP\cite{madsen2006boltztrap} code. 
Lattice points and their respective star functions 
are generated in the real space 
following point group operations of crystal symmetry. 
The corresponding translation vectors can be given as  
${\bf{R}} = u_1{\bf{a_1}} + u_2{\bf{a_2}} + u_3{\bf{a_3}}$, in which 
${\bf{a_1}}$, ${\bf{a_2}}$, ${\bf{a_3}}$ are 
related to the crystal's direct primitive vectors.
Such points are generated inside a sphere with a radius defined as  
$R^{\prime} = \sqrt[3]{3\cdot n_{{\bf{\bar{k}}}} \cdot n_s \cdot M \cdot \Omega/4\pi}$,   
in which $\Omega$ is the volume of the unit cell.
Consequently, $R^{\prime}$ determines the full extension of the real space and can be 
properly increased, for example, by increasing $M$, 
the number of star functions per ${\bf{k}}$-point.
In order to capture crystal anisotropy, 
the extension of the real space can be determined for each 
crystal direction, defining spheres for each crystallographic 
axis with the maximum radius given by $R_{max}(t) = INT(R^{\prime}\cdot\sqrt{{\bf{b_t}}\cdot{\bf{b_t}}}) + 1$, 
where ${\bf{b_t}}$ are the respective reciprocal primitive vectors, with $t=\{1,2,3\}$,   
and $INT(x)$ takes the largest integer number 
that does not exceed the magnitude of $x$. 

The star functions are ordered in such a way 
that the magnitude of $\bf{R}_m$ is nondecreasing as $m$ increases.
Thus, a 3D array containing all vectors 
are sorted considering their 
concentric radius, $r$, from the sphere center  
defined for each axis, and provided that 
all vectors, ${\bf{R}}$, have different star functions, $m$. 
The magnitude of each $\bf{R}_m$ vector is defined
through the metric tensor formalism.
For all $\bf{R}_m$ in the Bravais lattice, the reciprocal lattice is characterized
by a set of wavevectors ${\bf{k}}$, such that, $e^{2\pi i {\bf{k}}\cdot{\bf{R}}_m} = 1$.
Given ${\bf{R}}_m$ and ${\bf{k}}$ in the same direction, the
magnitude of the vector ${\bf{k}}$ in the reciprocal space is given by
$\left\vert {\bf{k}} \right\vert = (k_1u_1 + k_2u_2 + k_3u_3)/r = (n_{int}(1)+n_{int}(2)+n_{int}(3))/r$,
where $n_{int}(t) = 1,2,...,k_{max}(t)$
are integer numbers with $k_{max}(t)=2R_{max}(t)+1$.  
To determine all $\bf{k}$ vectors from $\bf{R}_m$, 
a 3D FFT is performed. 
In practice, $k_{max}(t)$ defines the number of data points on each dimension 
and should be carefully taken as the product of small primes in order 
to improve the efficiency of FFT. 

In fact, the FFT computational complexity is 
$\mathcal{O}(N\log{}N)$, where $N$ corresponds to the number of 
data points related to the product of FFT dimensions, 
namely $N = k_{max}(1)\times k_{max}(2)\times k_{max}(3)\approx 8R^{{\prime}^3}\cdot\sqrt{{\bf{b_1}}\cdot{\bf{b_1}}}\cdot\sqrt{{\bf{b_2}}\cdot{\bf{b_2}}}\cdot\sqrt{{\bf{b_3}}\cdot{\bf{b_3}}} \approx 6/\pi({n_{{\bf{\bar{k}}}}}\times n_s \times M)$. 
Consequently, the number of floating-point operations 
by using our approach is $\approx N_{f}^{{\bf{q^{\prime}}}}\times {n_{{\bf{\bar{k}}}}} n^3 + 
6/\pi({n_{{\bf{\bar{k}}}}}\times n_s \times M)\ln{(6/\pi({n_{{\bf{\bar{k}}}}}\times n_s \times M))}$. 
The first term comes from the 
first W-F interpolation by using $n_{{\bf{\bar{k}}}}$ irreducible points, while the second one 
comes from the symmetry-adapted PW interpolation. 
The gain in performance by using our method in comparison with single W-F  
calculation, to get approximately the same final homogeneous 
grid, can be given by $\approx 2 (n_s \times M)$, 
assuming $N_{f}^{\bf{k^{\prime}}} \approx N$ and $N_{f}^{{\bf{k^{\prime}}}} = N_{f}^{{\bf{q^{\prime}}}}$. 
Clearly, high symmetry systems allow greater computational savings, however 
the factor $M$, typically ranging from $5-60$ enables 
remarkably significant performance gain even for low symmetry systems.

In order to test our implementation, we carried out {\it{TP}} calculations for silicon.
We computed the imaginary part of the electron self-energy 
in the Fan-Migdal approximation, $\Im \Sigma$,
which gives the relaxation time due to {\it{e-ph}} scattering, and consequently,
thermoelectric {\it{TP}} using the BTE, as shown in Fig.~\ref{fig1}. Additionally,
we also studied phonon-assisted optical absorption for Si, as shown in Fig.~\ref{fig2}.
More details about these calculations can be found in the Supplemental Material (SM).
Since the first step in computing the double BZ integrals is
based on a W-F interpolation, our implementation
has been built on top of the Electron-Phonon Wannier (EPW)~\cite{ponce2016epw} code,
which is contained in the Quantum Espresso package~\cite{giannozzi2009quantum}.
We have modified the EPW code in order to
include the second PW interpolation, as described above, 
which we call Turbo-EPW (T-EPW).
In Fig.~\ref{fig3}(a) we show $\Im \Sigma$ for Si
at 300 K calculated using T-EPW in comparison with
other approaches, namely, previous DFT with linear
interpolation\cite{sun2012intrinsic,restrepo2009first,li2015electrical}, 
W-F interpolation with EPW\cite{qiu2015first}, and tight-binding calculations\cite{Rideau}.
Our results, calculated using $(100)^3{\bf{k^{\prime}}}$/$(100)^3{\bf{q^{\prime}}}$ grids,
are in good agreement with other W-F calculations using the EPW code directly on
$(30)^3{\bf{k^{\prime}}}/(60)^3{\bf{q^{\prime}}}$\cite{qiu2015first}
and $(100)^3{\bf{k^{\prime}}}/(40)^3{\bf{q^{\prime}}}$ grids, 
but with significantly reduced computational time.

In order to estimate the performance gain of our approach, in Fig.~{\ref{fig3}}(b)  
we show the computational time required to finalize the calculation
of $\Im \Sigma$ for different ${\bf{k^{\prime}}}$/${\bf{q^{\prime}}}$ grids, all using the same computational hardware.
The time required for calculations
based only on W-F interpolations (EPW) grows almost exponentially with increasing
${\bf{k^{\prime}}}$/${\bf{q^{\prime}}}$ density. By applying our method a drastic
reduction in the computational time is obtained, which is generally greater than two
orders of magnitude, as can be observed from the curves for the total time of T-EPW.
These curves also demonstrate a roughly exponential growth 
with ${\bf{k^{\prime}}}$/${\bf{q^{\prime}}}$ density, which is due to 
increasing the grid size in the first W-F interpolation,
independent of the number of initial irreducible points ${\bf{\bar{k}}}_l$. In these test calculations
we considered $n_{{\bf{\bar{k}}}} = 256, 1661, 5216$,
leading to regular meshes of ${\bf{{k^r}}} = (20)^3, (40)^3, (50)^3$.
The plateaus in the computational 
time for T-EPW are due to increasing the value of $M$ from $5$ to $60$,
while keeping fixed the grid of first W-F interpolation ($(100)^3 {\bf{q^{\prime}}}$ points).
As shown in Fig.~\ref{fig3}(b), using PW interpolation one can 
achieve much denser grids by increasing the value of $M$ with 
negligible increase in computational time.
As shown in the SM, the accompanying error due to the increase of $M$ to generate denser grids 
diminishes by increasing $n_{{\bf{{\bar{k}}}}}$; a solution that can also be used 
to minimize errors from possible kink structures derived from band crossings, which leads
to a Gibbs ringing in Fourier series analysis.

In summary, our method can be used to calculate efficiently {\it{el-ph}}-based ${\it{TP}}$.
The computational performance gain is remarkable, being $\approx 2 (n_s \times M)$ faster 
than state-of-the-art EPW calculations without loosing accuracy. 
It should be emphasized that this novel approach can also be used as an efficient
and stable numerical tool in order to calculate ubiquitous double BZ integrals,
and potentially extending to many further applications, for instance
phonon-assisted nonlinear optical properties,
superconducting critical temperature and its related thermodynamic
properties and electron-plasmon coupling ${\it{TP}}$ from first-principles.
Moreover, this method may allow previously impractical calculations 
and can serve as a starting point to explore
the effects of the vertex corrections to the Migdal
approximation as well as to address the {\it{e-ph}} coupling in complex
systems with many atoms in the unit cell. 
This last capability would be useful 
for the discovery of efficient materials for 
energy applications, such as high-performance thermoelectrics.

\begin{appendices}

\section{Details of the calculations for Si}

First, we compute the self-consistent potential and Kohn-Sham states on a
$12\times12\times12$ Monkhorst-Pack {\bf{k}}-point grid using DFT and lattice-dynamical properties with
DFPT\cite{baroni2001phonons} on a $3\times3\times3$ {\bf{q}}-point grid,
as implemented in the Quantum Espresso distribution\cite{giannozzi2009quantum}
using the Perdew-Burke-Ernzerhoff exchange-correlation functional\cite{perdew1996generalized}.
We used a full-relativistic norm-conserving
optimized Vanderbilt pseudopotential\cite{scherpelz2016implementation}.
The unit cell consists of Si in the diamond structure with
an experimental lattice parameter of $5.43~\si{\angstrom}$.
The {\it{e-ph}} matrix elements are first computed
on coarse grids, then they are determined in the significantly finer grids using
both Wannier-Fourier (W-F) interpolation only, through EPW code
and our dual interpolation method, Turbo-EPW (T-EPW).
Maximally localized Wannier functions\cite{marzari1997maximally} for the
wannierization procedure are obtained
from Wannier90.\cite{mostofi2008wannier90} Thus, Bloch-to-Wannier
rotation matrices and then Wannier-to-Bloch diagonalizer matrices are used
to interpolate {\it{el-ph}} matrix elements.

\section{Convergence analysis}

\begin{figure}
\centering
\includegraphics[width=0.48\textwidth]{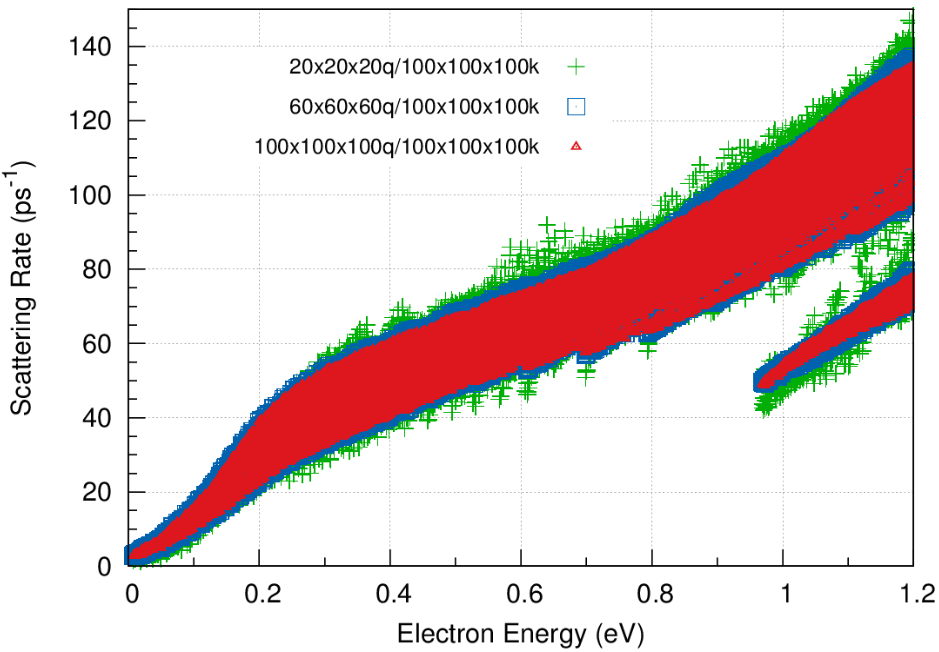}
\vspace*{6mm}\caption{\label{figureS1}
Convergence analysis of the scattering rate of Si at 300K
due to el-ph coupling in the Fan-Migdal approximation as a
function of the electron energy. The calculation has been
performed on different {\bf{q}} meshes, namely, $(20)^3$ (green crosses),
$(60)^3$ (blue squares), $(100)^3$ (red dots) {\bf{q}}-meshes in the
first W-F interpolation, while keeping fixed the number of irreducible
{\bf{k}} points in the second plane-waves interpolation,
leading to $(100)^3$ {\bf{k}}-mesh.}
\end{figure}

\begin{figure}
\centering
\includegraphics[width=0.48\textwidth]{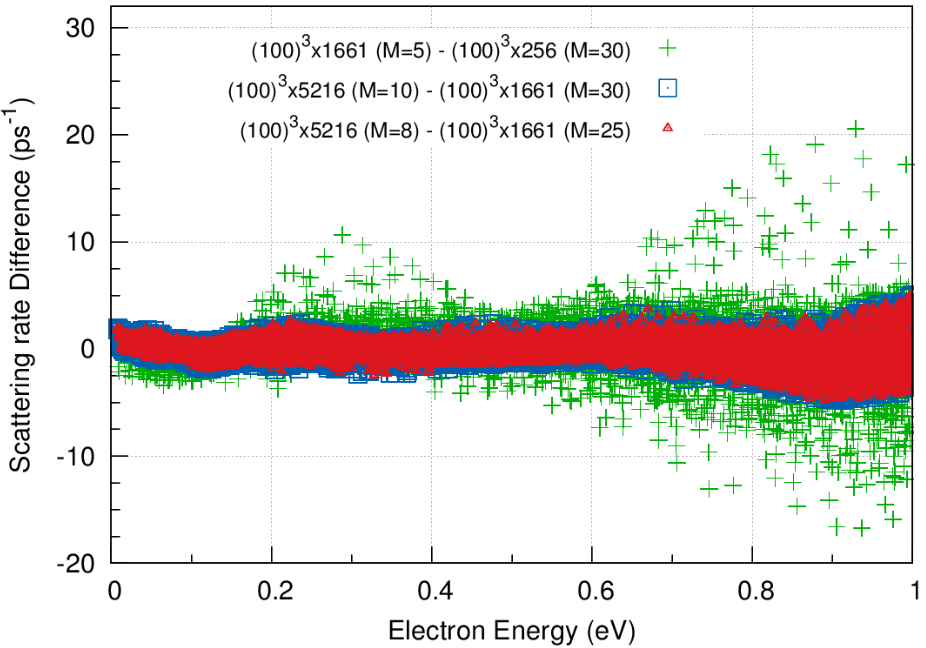}
\vspace*{6mm}\caption{\label{figureS2}
Difference between scattering rates computed by using
different parameters in the second plane-waves interpolation,
over equivalent meshes. Green crosses correspond to the difference
between scattering rates calculated over $(100)^3$ {\bf{q}}/ $1661$ {\bf{k}}
points expanded by using $M = 5$ to reach $(100)^3$ {\bf{k}}-mesh and
data from $(100)^3$ {\bf{q}}/ $256$ {\bf{k}} points expanded by using $M = 30$
to reach $(100)^3$ {\bf{k}}-mesh. Blue squares correspond to the
difference between scattering rates calculated over $(100)^3$ {\bf{q}}/
$5216$ {\bf{k}} points expanded by using $M = 10$ to reach $(180)^3$ {\bf{k}}-mesh
and data from $(100)^3$ {\bf{q}}/ $1661$ {\bf{k}} points expanded by
using $M = 30$ to reach $(180)^3$ {\bf{k}}-mesh. Red dots
correspond to the difference between scattering rates calculated over 
$(100)^3$ {\bf{q}}/ $5216$ {\bf{k}} points expanded by using $M = 8$
to reach $(160)^3$ {\bf{k}}-mesh and data from $(100)^3$
{\bf{q}}/ $1661$ {\bf{k}} points expanded by using $M = 25$ to reach
$(160)^3$ {\bf{k}}-mesh.}
\end{figure}

Fig.~\ref{figureS1} shows how the scattering rate aproaches
convergence by increasing the mesh size of W-F interpolation,
from $(20)^3$ to $(100)^{3}$ ${\bf{q}}$ points,
while keeping the number of irreducible points
fixed at $5216$ ${\bf{k}}$ points.
The second interpolation by star functions leads
into a converged grid with $(100)^3$ ${\bf{k}}$ points.
Fig.~\ref{figureS2} shows the difference in
scattering rates calculated by different approaches, namely,
different parameters in the second plane-waves interpolation (PWI),
over equivalent meshes.
The analysis shows that the accompanying error due to the increase in $M$
decreases by enlarging the number of irreducible points, $n_{{\bf{\bar{k}}}}$ (see main text).
Moreover, the difference between scattering rates calculated over $(100)^3$ ${\bf{q}}$/ $1661$ ${\bf{k}}$
points expanded by using $M = 5$ to reach $(100)^3$ ${\bf{k}}$-mesh and
data from $(100)^3$ ${\bf{q}}$/ $256$ ${\bf{k}}$ points expanded by using $M = 30$
to reach $(100)^3$ ${\bf{k}}$-mesh,
is within $\approx \pm 6\%$.
For the remaining, the difference between the data points is about
the same, within $\approx \pm 3\%$.

\section{Calculation of electron self-energy and thermoelectric properties}

The expression for the imaginary part of electronic self-energy due to
{\it{el-ph}} coupling in the
Fan-Migdal approximation can be derived from quantum field theory\cite{giustino2017electron}
and it is expressed as
\begin{equation}
\label{imag}
 \begin{split}
\Sigma^{\prime\prime}_{n,{\bf{k}}}(\omega,T)=\pi \sum_{m,\beta} \int_{BZ} \frac{d{\bf{q}}}{\Omega_{BZ}} |g_{mn,\beta}({\bf{k},{\bf{q}}})|^2 \\
       \times \Bigg[\left[n_{{\bf{q}}\beta}(T) + f_{m{\bf{k}}+{\bf{q}}}\right]\delta(\omega - (\epsilon_{m{\bf{k}}+{\bf{q}}} - \epsilon_F) + \omega_{{\bf{q}}\beta}) \\
        + [n_{{\bf{q}}\beta}(T) + 1 - f_{m{\bf{k}}+{\bf{q}}}]\delta(\omega - (\epsilon_{m{\bf{k}}+{\bf{q}}} - \epsilon_F) - \omega_{{\bf{q}}\beta})\Bigg]~,
\end{split}
\end{equation}
where $n_{{\bf{q}}\beta}(T)$ and $f_{m{\bf{k}}+{\bf{q}}}$ are the Bose-Einstein and the Fermi-Dirac distributions,
$\Omega_{BZ}$ is the BZ volume, $m$ and $n$ are the corresponding electronic states, while $\beta$ represents the
phonon branch, $\epsilon_{m{\bf{k}}+{\bf{q}}}$ are the electronic eigenenergies of the state $m{\bf{k}}+{\bf{q}}$ and
$\omega_{{\bf{q}}\beta}$ are the corresponding eigenfrequencies with wavevector ${\bf{q}}$ and phonon branch $\beta$.
Basically, from the first W-F interpolation we can get
$f({\bf{\bar{k}}}) = \Sigma^{\prime\prime}_{n,{\bf{\bar{k}}}}(\omega,T)$,
over the irreducible points,
which will be interpolated throughout the whole BZ by star functions,
resulting in $\Sigma^{\prime\prime}_{n,{\bf{k^{\prime}}}}(\omega,T)$ over denser
${\bf{k^{\prime}}}$ grids.

$\Sigma^{\prime\prime}$ is directly related to the scattering rate, that is, inversely
proportional to the relaxation time 
\begin{equation}
\label{tau}
\frac{1}{\tau_{n,{\bf{k}}}} = 2\Sigma^{\prime\prime}_{n,{\bf{k}}}(\omega = 0,T)~,
\end{equation}
which enters in kinetic transport equations.
Indeed, the kinetic coefficient tensors can be expressed through
\begin{equation}
\label{Lambda}
\Lambda^{(\alpha)}(\mu;T)= e{^2}\int\Xi(\epsilon,\mu,T)(\epsilon - \mu)^{\alpha}\left(-\frac{\partial f^{(0)}(\mu;\epsilon,T)}{\partial \epsilon}\right)d\epsilon~, 
\end{equation}
where $\mu$ is the chemical potential and $\Xi(\epsilon,\mu,T)$
is the transport distribution kernel given by
\begin{equation}
\label{Xi}
\Xi(\epsilon,\mu,T) = \int \sum_n{{\bf{v}}_{n,{\bf{k}}}\otimes{\bf{v}}_{n,{\bf{k}}}{\tau}_{n,{\bf{k}}}}(\mu,T)\delta(\epsilon - \epsilon_{n,{\bf{k}}})\frac{d{\bf{k}}}{8\pi^3}~, 
\end{equation}
with ${\bf{v}}_{n,{\bf{k}}}$ being the electron velocity.
From both experimental conditions of zero temperature gradient ($\nabla T = 0$)
and zero electric current, the kinetic coefficient tensors
can be identified with the electrical conductivity tensor, $\sigma = \Lambda^{(0)}$,
the Seebeck coefficient tensor, $S = (eT)^{-1}\Lambda^{(1)}/\Lambda^{(0)}$,
and the charge carrier contribution to thermal conductivity tensor,
$\kappa_e = (e^2T)^{-1} \left({\Lambda^{(1)}\cdot{\Lambda^{(0)}}^{-1}}\cdot{\Lambda^{(1)}} - \Lambda^{(2)}\right)$.
Consequently, once we have $\tau_{n,{\bf{k}}}$ from Eq.~(\ref{tau}),
we can compute the electrical conductivity, the Seebeck coefficient, Lorenz function and
the charge carrier contribution to thermal conductivity from the solution of Eq.~(\ref{Xi}) and~Eq.~(\ref{Lambda}).
Note that the both bandstructure and phonon dispersion have also been interpolated by the method
presented in the main text over the same grid as electron self-energy.
Indeed, we have implemented these equations on top of the BoltzTraP code,\cite{madsen2006boltztrap}
from which we can obtain these transport properties
directly from first principles.

\section{Scattering by ionized impurities}

For the calculation of thermoelectric transport properties of
$n$- and $p$-type Si polycrystals, we have also considered the scattering 
by ionized impurities.
Such scattering has been treated theoretically by
Brooks and Herring (B-H)\cite{brooks1955theory,chattopadhyay1981electron}
by considering a screened Coulomb potential, the
Born approximation for the evaluation of transition probabilities
and neglecting perturbation effects of the
impurities on the electron energy levels and wave functions.
In the B-H theory the electron is scattered independently
by dilute concentrations of ionized centers randomly distributed in the semiconductor.

The per-unit-time transition probability for the scattering of charge carriers
by ionized impurities can be given in the plane-wave approximation as
\begin{multline}
\label{U}
W({\bf{k}}|{\bf{{k}^{\prime}}}) = \frac{2\pi}{\hbar}\frac{N_i}{V} \\
\left\vert{\int U({\bf{r}}) \exp{\left[i({\bf{k}}-{\bf{{k}^{\prime}}};{\bf{r}})\right]}d{\bf{r}}}\right\vert^2 \delta(\epsilon({\bf{{k}^{\prime}}})-\epsilon({\bf{k}}))~,
\end{multline}
where $U({\bf{r}})$ is the scattering potential and $N_i$ is the ionized impurity concentration.

A long-range Coulomb field,
$U(\mat{r}) = e\phi(r) = \pm e^2/\zeta r$, with potential $\phi$ at a point $r$ of the crystal
is created by the presence of positive (donor) or negative (acceptor)
impurity ions, within a medium with dieletric constant $\zeta$.
The straightforward application of this field in Eq.~(\ref{U}) leads
into a logarithmic divergence, and hence, a screened Coulomb potential has to be considered.
From the B-H theory the potential can be expressed in a more rigorous form as
$\phi(r) = \pm e/{\zeta r} \left(\exp{\left(-{r}/{r_0}\right)}\right)$,
where $r_0$ is the radius of ion field screening defined by
\begin{equation}
\label{r0}
r{_0}^{-2}({\bf{k}}) = \frac{4\pi{e^2}}{\zeta{_0}}\int{-\frac{\partial f_0}{\partial \epsilon({\bf{k}})}}g(\epsilon)d\epsilon~, 
\end{equation}
where $f^{(0)}(\epsilon)$ is the equilibrium electron
distribution function, $\zeta_0$ is the static dielectric constant,
and $g(\epsilon)$ is the density of states,
calculated numerically on an energy grid
with spacing $d\epsilon$ sampled over $N_k$ ${\bf{k}}$-points 
\begin{equation}
\label{dos2}
g(\epsilon) = \int \sum_n \delta(\epsilon - \epsilon_{n,{\bf{k}}}) \frac{d{\bf{k}}}{8\pi^3} = \frac{1}{\Omega N_k}\sum_{n,{\bf{k}}} \frac{\delta(\epsilon - \epsilon_{n,{\bf{k}}})}{d\epsilon}~, 
\end{equation}
where $\Omega$ is the volume of the unit cell.

Within the relaxation time approximation for the
Boltzmann transport equations, the relaxation time for the scattering of the
charge carriers by ionized impurities can be expressed as
\begin{equation}
\label{tau22}
\tau_{imp}({\bf{k}}) = \frac{\hbar\zeta{_0}{^2}}{2{\pi}{e^4}{N_i}F_{imp}({\bf{k}})}{\bf{k}}^2 \left\vert\frac{\partial \epsilon({\bf{k}})}{\partial {\bf{k}}}\right\vert 
\end{equation}
where
\begin{equation}
F_{imp}({\bf{k}}) = ln(1+\eta) - \frac{\eta}{1+\eta}~, 
\end{equation}
is the screening function with $\eta = (2{\bf{k}}r_0)^2$.
Here, we interpolated $\left\vert\frac{\partial \epsilon({\bf{k}})}{\partial {\bf{k}}}\right\vert$
within Eq.~(\ref{tau22}) by using the plane-waves interpolation (see the main text) and
used Mathiessen's rule to consider both the scattering by ionized impurities and phonons.

\section{Calculation of phonon-assisted optical absorption}

To calculate the phonon-assisted absorption coefficient,
we use the Fermi's golden rule expression\cite{parravicini1975g,noffsinger2012}:
\begin{eqnarray}
\label{imag}
\alpha(\omega) = 2 \frac{4\pi^2 e^2}{\omega c n_r(\omega)}\frac{1}{\Omega} \frac{1}{N_{{\bf{k}}} N_{{\bf{q}}}} \sum_{\beta i j {\bf{k}} {\bf{q}}}|{\mat{\lambda}} \cdot ({\bf{S_1}} + {\bf{S_2}})|^2 \\
\times P \delta(\epsilon_{j,{\bf{k}}+{\bf{q}}} - \epsilon_{i,{\bf{k}}} - \hbar\omega - \pm \hbar\omega_{\beta {\bf{q}}}) \nonumber ~,
\end{eqnarray}

with

\begin{equation} 
{\bf{S_1}}({\bf{k}} {\bf{q}}) = \sum_m \frac{{\bf{v}}_{im}({\bf{k}}) g_{m j,\beta}({\bf{k}},{\bf{q}})}{\epsilon_{m,{\bf{k}}} - \epsilon_{i {\bf{k}}} - \hbar\omega + i\Gamma_{m,{\bf{k}}}}~,
\end{equation}

\begin{equation}
{\bf{S_2}}({\bf{k}} {\bf{q}}) = \sum_m \frac{g_{i m,\beta}({\bf{k}},{\bf{q}}) {\bf{v}}_{mj}({\bf{k}} + {\bf{q}})}{\epsilon_{m,{\bf{k}} + {\bf{q}}} - \epsilon_{i {\bf{k}}} \pm \hbar\omega_{\beta {\bf{q}}} + i\Gamma_{m,{\bf{k}}+{\bf{q}}}} ~,
\end{equation}

and

\begin{equation}
P = (n_{\beta {\bf{q}}} + \frac{1}{2} \pm \frac{1}{2})(f_{i {\bf{k}}} - f_{j, {\bf{k}}+{\bf{q}}})~.
\end{equation}

In these equations, $\omega$ is the photon frequency,
$c$ is the speed of light, $n_r$ is the refractive index
(for silicon we used $n_r = 3.4$), and $\mat{\lambda}$ is
the photon polarization. ${\bf{S_1}}$ and ${\bf{S_2}}$ are
the two possible ways for the indirect absorption process, while
$P$ is related to the carrier and phonon statistics.
For calculations of Si, the DFT band gap and all
conduction bands have been shifted up by $0.7$\si{\eV}
to simulate experimental gap.
Our calculations have been carried out over $(60)^3${\bf{q}}/$(40)^3${\bf{k}}
meshes. It took $\approx 36$ minutes to perform the calculation by using
our dual interpolation method on $8$
CPU cores at the Odyssey cluster (Harvard University).

\end{appendices}

\setcounter{secnumdepth}{0}
\section{Acknowledgments} 
The authors thank the Harvard FAS Research Computing facility and the Brazilian
CCJDR-IFGW-UNICAMP for computational resources.
A.S.C. and A.A. gratefully acknowledge financial support from the 
Brazilian agency FAPESP under Grants No.2015/26434-2, 
No.2016/23891-6, No.2017/26105-4, No.2018/01274-0 and No.2019/26088-8.
A.S.C. also acknowledges the kind hospitality of SEAS-Harvard University.

\bibliographystyle{apsrev4-1} 
\bibliography{Ref.bib}

\begin{thebibliography}{34}%
\makeatletter
\providecommand \@ifxundefined [1]{%
 \@ifx{#1\undefined}
}%
\providecommand \@ifnum [1]{%
 \ifnum #1\expandafter \@firstoftwo
 \else \expandafter \@secondoftwo
 \fi
}%
\providecommand \@ifx [1]{%
 \ifx #1\expandafter \@firstoftwo
 \else \expandafter \@secondoftwo
 \fi
}%
\providecommand \natexlab [1]{#1}%
\providecommand \enquote  [1]{``#1''}%
\providecommand \bibnamefont  [1]{#1}%
\providecommand \bibfnamefont [1]{#1}%
\providecommand \citenamefont [1]{#1}%
\providecommand \href@noop [0]{\@secondoftwo}%
\providecommand \href [0]{\begingroup \@sanitize@url \@href}%
\providecommand \@href[1]{\@@startlink{#1}\@@href}%
\providecommand \@@href[1]{\endgroup#1\@@endlink}%
\providecommand \@sanitize@url [0]{\catcode `\\12\catcode `\$12\catcode
  `\&12\catcode `\#12\catcode `\^12\catcode `\_12\catcode `\%12\relax}%
\providecommand \@@startlink[1]{}%
\providecommand \@@endlink[0]{}%
\providecommand \url  [0]{\begingroup\@sanitize@url \@url }%
\providecommand \@url [1]{\endgroup\@href {#1}{\urlprefix }}%
\providecommand \urlprefix  [0]{URL }%
\providecommand \Eprint [0]{\href }%
\providecommand \doibase [0]{http://dx.doi.org/}%
\providecommand \selectlanguage [0]{\@gobble}%
\providecommand \bibinfo  [0]{\@secondoftwo}%
\providecommand \bibfield  [0]{\@secondoftwo}%
\providecommand \translation [1]{[#1]}%
\providecommand \BibitemOpen [0]{}%
\providecommand \bibitemStop [0]{}%
\providecommand \bibitemNoStop [0]{.\EOS\space}%
\providecommand \EOS [0]{\spacefactor3000\relax}%
\providecommand \BibitemShut  [1]{\csname bibitem#1\endcsname}%
\let\auto@bib@innerbib\@empty
\bibitem [{\citenamefont {Giustino}(2017)}]{giustino2017electron}%
  \BibitemOpen
  \bibfield  {author} {\bibinfo {author} {\bibfnamefont {F.}~\bibnamefont
  {Giustino}},\ }\href@noop {} {\bibfield  {journal} {\bibinfo  {journal}
  {Reviews of Modern Physics}\ }\textbf {\bibinfo {volume} {89}},\ \bibinfo
  {pages} {015003} (\bibinfo {year} {2017})}\BibitemShut {NoStop}%
\bibitem [{\citenamefont {Marini}(2008)}]{marini2008ab}%
  \BibitemOpen
  \bibfield  {author} {\bibinfo {author} {\bibfnamefont {A.}~\bibnamefont
  {Marini}},\ }\href@noop {} {\bibfield  {journal} {\bibinfo  {journal}
  {Physical Review Letters}\ }\textbf {\bibinfo {volume} {101}},\ \bibinfo
  {pages} {106405} (\bibinfo {year} {2008})}\BibitemShut {NoStop}%
\bibitem [{\citenamefont {Park}\ \emph {et~al.}(2007)\citenamefont {Park},
  \citenamefont {Giustino}, \citenamefont {Cohen},\ and\ \citenamefont
  {Louie}}]{park2007velocity}%
  \BibitemOpen
  \bibfield  {author} {\bibinfo {author} {\bibfnamefont {C.-H.}\ \bibnamefont
  {Park}}, \bibinfo {author} {\bibfnamefont {F.}~\bibnamefont {Giustino}},
  \bibinfo {author} {\bibfnamefont {M.~L.}\ \bibnamefont {Cohen}}, \ and\
  \bibinfo {author} {\bibfnamefont {S.~G.}\ \bibnamefont {Louie}},\ }\href@noop
  {} {\bibfield  {journal} {\bibinfo  {journal} {Physical Review Letters}\
  }\textbf {\bibinfo {volume} {99}},\ \bibinfo {pages} {086804} (\bibinfo
  {year} {2007})}\BibitemShut {NoStop}%
\bibitem [{\citenamefont {Restrepo}\ \emph {et~al.}(2009)\citenamefont
  {Restrepo}, \citenamefont {Varga},\ and\ \citenamefont
  {Pantelides}}]{restrepo2009first}%
  \BibitemOpen
  \bibfield  {author} {\bibinfo {author} {\bibfnamefont {O.}~\bibnamefont
  {Restrepo}}, \bibinfo {author} {\bibfnamefont {K.}~\bibnamefont {Varga}}, \
  and\ \bibinfo {author} {\bibfnamefont {S.}~\bibnamefont {Pantelides}},\
  }\href@noop {} {\bibfield  {journal} {\bibinfo  {journal} {Applied Physics
  Letters}\ }\textbf {\bibinfo {volume} {94}},\ \bibinfo {pages} {212103}
  (\bibinfo {year} {2009})}\BibitemShut {NoStop}%
\bibitem [{\citenamefont {Margine}\ and\ \citenamefont
  {Giustino}(2014)}]{margine2014two}%
  \BibitemOpen
  \bibfield  {author} {\bibinfo {author} {\bibfnamefont {E.}~\bibnamefont
  {Margine}}\ and\ \bibinfo {author} {\bibfnamefont {F.}~\bibnamefont
  {Giustino}},\ }\href@noop {} {\bibfield  {journal} {\bibinfo  {journal}
  {Physical Review B}\ }\textbf {\bibinfo {volume} {90}},\ \bibinfo {pages}
  {014518} (\bibinfo {year} {2014})}\BibitemShut {NoStop}%
\bibitem [{\citenamefont {Fiorentini}\ and\ \citenamefont
  {Bonini}(2016)}]{fiorentini2016thermoelectric}%
  \BibitemOpen
  \bibfield  {author} {\bibinfo {author} {\bibfnamefont {M.}~\bibnamefont
  {Fiorentini}}\ and\ \bibinfo {author} {\bibfnamefont {N.}~\bibnamefont
  {Bonini}},\ }\href@noop {} {\bibfield  {journal} {\bibinfo  {journal}
  {Physical Review B}\ }\textbf {\bibinfo {volume} {94}},\ \bibinfo {pages}
  {085204} (\bibinfo {year} {2016})}\BibitemShut {NoStop}%
\bibitem [{\citenamefont {Wang}\ \emph {et~al.}(2011)\citenamefont {Wang},
  \citenamefont {Wang}, \citenamefont {Obukhov}, \citenamefont {Vast},
  \citenamefont {Sjakste}, \citenamefont {Tyuterev},\ and\ \citenamefont
  {Mingo}}]{wang2011thermoelectric}%
  \BibitemOpen
  \bibfield  {author} {\bibinfo {author} {\bibfnamefont {Z.}~\bibnamefont
  {Wang}}, \bibinfo {author} {\bibfnamefont {S.}~\bibnamefont {Wang}}, \bibinfo
  {author} {\bibfnamefont {S.}~\bibnamefont {Obukhov}}, \bibinfo {author}
  {\bibfnamefont {N.}~\bibnamefont {Vast}}, \bibinfo {author} {\bibfnamefont
  {J.}~\bibnamefont {Sjakste}}, \bibinfo {author} {\bibfnamefont
  {V.}~\bibnamefont {Tyuterev}}, \ and\ \bibinfo {author} {\bibfnamefont
  {N.}~\bibnamefont {Mingo}},\ }\href@noop {} {\bibfield  {journal} {\bibinfo
  {journal} {Physical Review B}\ }\textbf {\bibinfo {volume} {83}},\ \bibinfo
  {pages} {205208} (\bibinfo {year} {2011})}\BibitemShut {NoStop}%
\bibitem [{\citenamefont {Noffsinger}\ \emph {et~al.}(2012)\citenamefont
  {Noffsinger}, \citenamefont {Kioupakis}, \citenamefont {Van~de Walle},
  \citenamefont {Louie},\ and\ \citenamefont {Cohen}}]{noffsinger2012}%
  \BibitemOpen
  \bibfield  {author} {\bibinfo {author} {\bibfnamefont {J.}~\bibnamefont
  {Noffsinger}}, \bibinfo {author} {\bibfnamefont {E.}~\bibnamefont
  {Kioupakis}}, \bibinfo {author} {\bibfnamefont {C.~G.}\ \bibnamefont {Van~de
  Walle}}, \bibinfo {author} {\bibfnamefont {S.~G.}\ \bibnamefont {Louie}}, \
  and\ \bibinfo {author} {\bibfnamefont {M.~L.}\ \bibnamefont {Cohen}},\
  }\href@noop {} {\bibfield  {journal} {\bibinfo  {journal} {Physical Review
  Letters}\ }\textbf {\bibinfo {volume} {108}},\ \bibinfo {pages} {167402}
  (\bibinfo {year} {2012})}\BibitemShut {NoStop}%
\bibitem [{\citenamefont {Boukai}\ \emph {et~al.}(2011)\citenamefont {Boukai},
  \citenamefont {Bunimovich}, \citenamefont {Tahir-Kheli}, \citenamefont {Yu},
  \citenamefont {Goddard~III},\ and\ \citenamefont
  {Heath}}]{boukai2011silicon}%
  \BibitemOpen
  \bibfield  {author} {\bibinfo {author} {\bibfnamefont {A.~I.}\ \bibnamefont
  {Boukai}}, \bibinfo {author} {\bibfnamefont {Y.}~\bibnamefont {Bunimovich}},
  \bibinfo {author} {\bibfnamefont {J.}~\bibnamefont {Tahir-Kheli}}, \bibinfo
  {author} {\bibfnamefont {J.-K.}\ \bibnamefont {Yu}}, \bibinfo {author}
  {\bibfnamefont {W.~A.}\ \bibnamefont {Goddard~III}}, \ and\ \bibinfo {author}
  {\bibfnamefont {J.~R.}\ \bibnamefont {Heath}},\ }in\ \href@noop {} {\emph
  {\bibinfo {booktitle} {Materials For Sustainable Energy: A Collection of
  Peer-Reviewed Research and Review Articles from Nature Publishing Group}}}\
  (\bibinfo  {publisher} {World Scientific},\ \bibinfo {year} {2011})\ pp.\
  \bibinfo {pages} {116--119}\BibitemShut {NoStop}%
\bibitem [{\citenamefont {Strasser}\ \emph {et~al.}(2004)\citenamefont
  {Strasser}, \citenamefont {Aigner}, \citenamefont {Lauterbach}, \citenamefont
  {Sturm}, \citenamefont {Franosch},\ and\ \citenamefont
  {Wachutka}}]{strasser2004micromachined}%
  \BibitemOpen
  \bibfield  {author} {\bibinfo {author} {\bibfnamefont {M.}~\bibnamefont
  {Strasser}}, \bibinfo {author} {\bibfnamefont {R.}~\bibnamefont {Aigner}},
  \bibinfo {author} {\bibfnamefont {C.}~\bibnamefont {Lauterbach}}, \bibinfo
  {author} {\bibfnamefont {T.}~\bibnamefont {Sturm}}, \bibinfo {author}
  {\bibfnamefont {M.}~\bibnamefont {Franosch}}, \ and\ \bibinfo {author}
  {\bibfnamefont {G.}~\bibnamefont {Wachutka}},\ }\href@noop {} {\bibfield
  {journal} {\bibinfo  {journal} {Sensors and Actuators A: Physical}\ }\textbf
  {\bibinfo {volume} {114}},\ \bibinfo {pages} {362} (\bibinfo {year}
  {2004})}\BibitemShut {NoStop}%
\bibitem [{\citenamefont {Baroni}\ \emph {et~al.}(2001)\citenamefont {Baroni},
  \citenamefont {De~Gironcoli}, \citenamefont {Dal~Corso},\ and\ \citenamefont
  {Giannozzi}}]{baroni2001phonons}%
  \BibitemOpen
  \bibfield  {author} {\bibinfo {author} {\bibfnamefont {S.}~\bibnamefont
  {Baroni}}, \bibinfo {author} {\bibfnamefont {S.}~\bibnamefont
  {De~Gironcoli}}, \bibinfo {author} {\bibfnamefont {A.}~\bibnamefont
  {Dal~Corso}}, \ and\ \bibinfo {author} {\bibfnamefont {P.}~\bibnamefont
  {Giannozzi}},\ }\href@noop {} {\bibfield  {journal} {\bibinfo  {journal}
  {Reviews of Modern Physics}\ }\textbf {\bibinfo {volume} {73}},\ \bibinfo
  {pages} {515} (\bibinfo {year} {2001})}\BibitemShut {NoStop}%
\bibitem [{\citenamefont {Li}(2015)}]{li2015electrical}%
  \BibitemOpen
  \bibfield  {author} {\bibinfo {author} {\bibfnamefont {W.}~\bibnamefont
  {Li}},\ }\href@noop {} {\bibfield  {journal} {\bibinfo  {journal} {Physical
  Review B}\ }\textbf {\bibinfo {volume} {92}},\ \bibinfo {pages} {075405}
  (\bibinfo {year} {2015})}\BibitemShut {NoStop}%
\bibitem [{\citenamefont {Giustino}\ \emph {et~al.}(2007)\citenamefont
  {Giustino}, \citenamefont {Cohen},\ and\ \citenamefont
  {Louie}}]{giustino2007electron}%
  \BibitemOpen
  \bibfield  {author} {\bibinfo {author} {\bibfnamefont {F.}~\bibnamefont
  {Giustino}}, \bibinfo {author} {\bibfnamefont {M.~L.}\ \bibnamefont {Cohen}},
  \ and\ \bibinfo {author} {\bibfnamefont {S.~G.}\ \bibnamefont {Louie}},\
  }\href@noop {} {\bibfield  {journal} {\bibinfo  {journal} {Physical Review
  B}\ }\textbf {\bibinfo {volume} {76}},\ \bibinfo {pages} {165108} (\bibinfo
  {year} {2007})}\BibitemShut {NoStop}%
\bibitem [{\citenamefont {Marzari}\ and\ \citenamefont
  {Vanderbilt}(1997)}]{marzari1997maximally}%
  \BibitemOpen
  \bibfield  {author} {\bibinfo {author} {\bibfnamefont {N.}~\bibnamefont
  {Marzari}}\ and\ \bibinfo {author} {\bibfnamefont {D.}~\bibnamefont
  {Vanderbilt}},\ }\href@noop {} {\bibfield  {journal} {\bibinfo  {journal}
  {Physical Review B}\ }\textbf {\bibinfo {volume} {56}},\ \bibinfo {pages}
  {12847} (\bibinfo {year} {1997})}\BibitemShut {NoStop}%
\bibitem [{\citenamefont {Braunstein}\ \emph {et~al.}(1958)\citenamefont
  {Braunstein}, \citenamefont {Moore},\ and\ \citenamefont
  {Herman}}]{braunstein1958intrinsic}%
  \BibitemOpen
  \bibfield  {author} {\bibinfo {author} {\bibfnamefont {R.}~\bibnamefont
  {Braunstein}}, \bibinfo {author} {\bibfnamefont {A.~R.}\ \bibnamefont
  {Moore}}, \ and\ \bibinfo {author} {\bibfnamefont {F.}~\bibnamefont
  {Herman}},\ }\href@noop {} {\bibfield  {journal} {\bibinfo  {journal}
  {Physical Review}\ }\textbf {\bibinfo {volume} {109}},\ \bibinfo {pages}
  {695} (\bibinfo {year} {1958})}\BibitemShut {NoStop}%
\bibitem [{\citenamefont {Fetter}\ and\ \citenamefont
  {Walecka}(2012)}]{fetter2012quantum}%
  \BibitemOpen
  \bibfield  {author} {\bibinfo {author} {\bibfnamefont {A.~L.}\ \bibnamefont
  {Fetter}}\ and\ \bibinfo {author} {\bibfnamefont {J.~D.}\ \bibnamefont
  {Walecka}},\ }\href@noop {} {\emph {\bibinfo {title} {Quantum theory of
  many-particle systems}}}\ (\bibinfo  {publisher} {Courier Corporation},\
  \bibinfo {year} {2012})\BibitemShut {NoStop}%
\bibitem [{\citenamefont {Pick}\ \emph {et~al.}(1970)\citenamefont {Pick},
  \citenamefont {Cohen},\ and\ \citenamefont {Martin}}]{pick1970microscopic}%
  \BibitemOpen
  \bibfield  {author} {\bibinfo {author} {\bibfnamefont {R.~M.}\ \bibnamefont
  {Pick}}, \bibinfo {author} {\bibfnamefont {M.~H.}\ \bibnamefont {Cohen}}, \
  and\ \bibinfo {author} {\bibfnamefont {R.~M.}\ \bibnamefont {Martin}},\
  }\href@noop {} {\bibfield  {journal} {\bibinfo  {journal} {Physical Review
  B}\ }\textbf {\bibinfo {volume} {1}},\ \bibinfo {pages} {910} (\bibinfo
  {year} {1970})}\BibitemShut {NoStop}%
\bibitem [{\citenamefont {Ponc{\'e}}\ \emph {et~al.}(2014)\citenamefont
  {Ponc{\'e}}, \citenamefont {Antonius}, \citenamefont {Boulanger},
  \citenamefont {Cannuccia}, \citenamefont {Marini}, \citenamefont
  {C{\^o}t{\'e}},\ and\ \citenamefont {Gonze}}]{ponce2014verification}%
  \BibitemOpen
  \bibfield  {author} {\bibinfo {author} {\bibfnamefont {S.}~\bibnamefont
  {Ponc{\'e}}}, \bibinfo {author} {\bibfnamefont {G.}~\bibnamefont {Antonius}},
  \bibinfo {author} {\bibfnamefont {P.}~\bibnamefont {Boulanger}}, \bibinfo
  {author} {\bibfnamefont {E.}~\bibnamefont {Cannuccia}}, \bibinfo {author}
  {\bibfnamefont {A.}~\bibnamefont {Marini}}, \bibinfo {author} {\bibfnamefont
  {M.}~\bibnamefont {C{\^o}t{\'e}}}, \ and\ \bibinfo {author} {\bibfnamefont
  {X.}~\bibnamefont {Gonze}},\ }\href@noop {} {\bibfield  {journal} {\bibinfo
  {journal} {Computational Materials Science}\ }\textbf {\bibinfo {volume}
  {83}},\ \bibinfo {pages} {341} (\bibinfo {year} {2014})}\BibitemShut
  {NoStop}%
\bibitem [{\citenamefont {Chadi}\ and\ \citenamefont
  {Cohen}(1973)}]{chadi1973special}%
  \BibitemOpen
  \bibfield  {author} {\bibinfo {author} {\bibfnamefont {D.}~\bibnamefont
  {Chadi}}\ and\ \bibinfo {author} {\bibfnamefont {M.~L.}\ \bibnamefont
  {Cohen}},\ }\href@noop {} {\bibfield  {journal} {\bibinfo  {journal}
  {Physical Review B}\ }\textbf {\bibinfo {volume} {8}},\ \bibinfo {pages}
  {5747} (\bibinfo {year} {1973})}\BibitemShut {NoStop}%
\bibitem [{\citenamefont {Shankland}(1971)}]{Shankland}%
  \BibitemOpen
  \bibfield  {author} {\bibinfo {author} {\bibfnamefont {D.~G.}\ \bibnamefont
  {Shankland}},\ }in\ \href@noop {} {\emph {\bibinfo {booktitle} {Computational
  Methods in Band Theory}}}\ (\bibinfo  {publisher} {Plenum},\ \bibinfo
  {address} {New York},\ \bibinfo {year} {1971})\ p.\ \bibinfo {pages}
  {362}\BibitemShut {NoStop}%
\bibitem [{\citenamefont {Koelling}\ and\ \citenamefont
  {Wood}(1986)}]{koelling1986interpolation}%
  \BibitemOpen
  \bibfield  {author} {\bibinfo {author} {\bibfnamefont {D.}~\bibnamefont
  {Koelling}}\ and\ \bibinfo {author} {\bibfnamefont {J.}~\bibnamefont
  {Wood}},\ }\href@noop {} {\bibfield  {journal} {\bibinfo  {journal} {Journal
  of Computational Physics}\ }\textbf {\bibinfo {volume} {67}},\ \bibinfo
  {pages} {253} (\bibinfo {year} {1986})}\BibitemShut {NoStop}%
\bibitem [{\citenamefont {Pickett}\ \emph {et~al.}(1988)\citenamefont
  {Pickett}, \citenamefont {Krakauer},\ and\ \citenamefont {Allen}}]{pickett}%
  \BibitemOpen
  \bibfield  {author} {\bibinfo {author} {\bibfnamefont {W.~E.}\ \bibnamefont
  {Pickett}}, \bibinfo {author} {\bibfnamefont {H.}~\bibnamefont {Krakauer}}, \
  and\ \bibinfo {author} {\bibfnamefont {P.~B.}\ \bibnamefont {Allen}},\
  }\href@noop {} {\bibfield  {journal} {\bibinfo  {journal} {Physical Review
  B}\ }\textbf {\bibinfo {volume} {38}},\ \bibinfo {pages} {2721} (\bibinfo
  {year} {1988})}\BibitemShut {NoStop}%
\bibitem [{\citenamefont {Sun}\ \emph {et~al.}(2012)\citenamefont {Sun},
  \citenamefont {Boggs},\ and\ \citenamefont {Ramprasad}}]{sun2012intrinsic}%
  \BibitemOpen
  \bibfield  {author} {\bibinfo {author} {\bibfnamefont {Y.}~\bibnamefont
  {Sun}}, \bibinfo {author} {\bibfnamefont {S.}~\bibnamefont {Boggs}}, \ and\
  \bibinfo {author} {\bibfnamefont {R.}~\bibnamefont {Ramprasad}},\ }\href@noop
  {} {\bibfield  {journal} {\bibinfo  {journal} {Applied Physics Letters}\
  }\textbf {\bibinfo {volume} {101}},\ \bibinfo {pages} {132906} (\bibinfo
  {year} {2012})}\BibitemShut {NoStop}%
\bibitem [{\citenamefont {Qiu}\ \emph {et~al.}(2015)\citenamefont {Qiu},
  \citenamefont {Tian}, \citenamefont {Vallabhaneni}, \citenamefont {Liao},
  \citenamefont {Mendoza}, \citenamefont {Restrepo}, \citenamefont {Ruan},\
  and\ \citenamefont {Chen}}]{qiu2015first}%
  \BibitemOpen
  \bibfield  {author} {\bibinfo {author} {\bibfnamefont {B.}~\bibnamefont
  {Qiu}}, \bibinfo {author} {\bibfnamefont {Z.}~\bibnamefont {Tian}}, \bibinfo
  {author} {\bibfnamefont {A.}~\bibnamefont {Vallabhaneni}}, \bibinfo {author}
  {\bibfnamefont {B.}~\bibnamefont {Liao}}, \bibinfo {author} {\bibfnamefont
  {J.~M.}\ \bibnamefont {Mendoza}}, \bibinfo {author} {\bibfnamefont {O.~D.}\
  \bibnamefont {Restrepo}}, \bibinfo {author} {\bibfnamefont {X.}~\bibnamefont
  {Ruan}}, \ and\ \bibinfo {author} {\bibfnamefont {G.}~\bibnamefont {Chen}},\
  }\href@noop {} {\bibfield  {journal} {\bibinfo  {journal} {EPL (Europhysics
  Letters)}\ }\textbf {\bibinfo {volume} {109}},\ \bibinfo {pages} {57006}
  (\bibinfo {year} {2015})},\ \bibinfo {note} {arXiv:1409.4862
  (2014)}\BibitemShut {NoStop}%
\bibitem [{\citenamefont {Rideau}\ \emph {et~al.}(2011)\citenamefont {Rideau},
  \citenamefont {Zhang}, \citenamefont {Niquet}, \citenamefont {Delerue},
  \citenamefont {Tavernier},\ and\ \citenamefont {Jaouen}}]{Rideau}%
  \BibitemOpen
  \bibfield  {author} {\bibinfo {author} {\bibfnamefont {D.}~\bibnamefont
  {Rideau}}, \bibinfo {author} {\bibfnamefont {W.}~\bibnamefont {Zhang}},
  \bibinfo {author} {\bibfnamefont {Y.}~\bibnamefont {Niquet}}, \bibinfo
  {author} {\bibfnamefont {C.}~\bibnamefont {Delerue}}, \bibinfo {author}
  {\bibfnamefont {C.}~\bibnamefont {Tavernier}}, \ and\ \bibinfo {author}
  {\bibfnamefont {H.}~\bibnamefont {Jaouen}},\ }in\ \href@noop {} {\emph
  {\bibinfo {booktitle} {2011 International Conference on Simulation of
  Semiconductor Processes and Devices (SISPAD)}}}\ (\bibinfo  {publisher}
  {IEEE},\ \bibinfo {address} {New York},\ \bibinfo {year} {2011})\ pp.\
  \bibinfo {pages} {47--50}\BibitemShut {NoStop}%
\bibitem [{\citenamefont {Madsen}\ and\ \citenamefont
  {Singh}(2006)}]{madsen2006boltztrap}%
  \BibitemOpen
  \bibfield  {author} {\bibinfo {author} {\bibfnamefont {G.~K.}\ \bibnamefont
  {Madsen}}\ and\ \bibinfo {author} {\bibfnamefont {D.~J.}\ \bibnamefont
  {Singh}},\ }\href@noop {} {\bibfield  {journal} {\bibinfo  {journal}
  {Computer Physics Communications}\ }\textbf {\bibinfo {volume} {175}},\
  \bibinfo {pages} {67} (\bibinfo {year} {2006})}\BibitemShut {NoStop}%
\bibitem [{\citenamefont {Ponc{\'e}}\ \emph {et~al.}(2016)\citenamefont
  {Ponc{\'e}}, \citenamefont {Margine}, \citenamefont {Verdi},\ and\
  \citenamefont {Giustino}}]{ponce2016epw}%
  \BibitemOpen
  \bibfield  {author} {\bibinfo {author} {\bibfnamefont {S.}~\bibnamefont
  {Ponc{\'e}}}, \bibinfo {author} {\bibfnamefont {E.~R.}\ \bibnamefont
  {Margine}}, \bibinfo {author} {\bibfnamefont {C.}~\bibnamefont {Verdi}}, \
  and\ \bibinfo {author} {\bibfnamefont {F.}~\bibnamefont {Giustino}},\
  }\href@noop {} {\bibfield  {journal} {\bibinfo  {journal} {Computer Physics
  Communications}\ }\textbf {\bibinfo {volume} {209}},\ \bibinfo {pages} {116}
  (\bibinfo {year} {2016})}\BibitemShut {NoStop}%
\bibitem [{\citenamefont {Giannozzi}\ \emph {et~al.}(2009)\citenamefont
  {Giannozzi} \emph {et~al.}}]{giannozzi2009quantum}%
  \BibitemOpen
  \bibfield  {author} {\bibinfo {author} {\bibfnamefont {P.}~\bibnamefont
  {Giannozzi}} \emph {et~al.},\ }\href@noop {} {\bibfield  {journal} {\bibinfo
  {journal} {Journal of Physics: Condensed Matter}\ }\textbf {\bibinfo {volume}
  {21}},\ \bibinfo {pages} {395502} (\bibinfo {year} {2009})}\BibitemShut
  {NoStop}%
\bibitem [{\citenamefont {Perdew}\ \emph {et~al.}(1996)\citenamefont {Perdew},
  \citenamefont {Burke},\ and\ \citenamefont
  {Ernzerhof}}]{perdew1996generalized}%
  \BibitemOpen
  \bibfield  {author} {\bibinfo {author} {\bibfnamefont {J.~P.}\ \bibnamefont
  {Perdew}}, \bibinfo {author} {\bibfnamefont {K.}~\bibnamefont {Burke}}, \
  and\ \bibinfo {author} {\bibfnamefont {M.}~\bibnamefont {Ernzerhof}},\
  }\href@noop {} {\bibfield  {journal} {\bibinfo  {journal} {Physical Review
  Letters}\ }\textbf {\bibinfo {volume} {77}},\ \bibinfo {pages} {3865}
  (\bibinfo {year} {1996})}\BibitemShut {NoStop}%
\bibitem [{\citenamefont {Scherpelz}\ \emph {et~al.}(2016)\citenamefont
  {Scherpelz}, \citenamefont {Govoni}, \citenamefont {Hamada},\ and\
  \citenamefont {Galli}}]{scherpelz2016implementation}%
  \BibitemOpen
  \bibfield  {author} {\bibinfo {author} {\bibfnamefont {P.}~\bibnamefont
  {Scherpelz}}, \bibinfo {author} {\bibfnamefont {M.}~\bibnamefont {Govoni}},
  \bibinfo {author} {\bibfnamefont {I.}~\bibnamefont {Hamada}}, \ and\ \bibinfo
  {author} {\bibfnamefont {G.}~\bibnamefont {Galli}},\ }\href@noop {}
  {\bibfield  {journal} {\bibinfo  {journal} {Journal of chemical theory and
  computation}\ }\textbf {\bibinfo {volume} {12}},\ \bibinfo {pages} {3523}
  (\bibinfo {year} {2016})}\BibitemShut {NoStop}%
\bibitem [{\citenamefont {Mostofi}\ \emph {et~al.}(2008)\citenamefont
  {Mostofi}, \citenamefont {Yates}, \citenamefont {Lee}, \citenamefont {Souza},
  \citenamefont {Vanderbilt},\ and\ \citenamefont
  {Marzari}}]{mostofi2008wannier90}%
  \BibitemOpen
  \bibfield  {author} {\bibinfo {author} {\bibfnamefont {A.~A.}\ \bibnamefont
  {Mostofi}}, \bibinfo {author} {\bibfnamefont {J.~R.}\ \bibnamefont {Yates}},
  \bibinfo {author} {\bibfnamefont {Y.-S.}\ \bibnamefont {Lee}}, \bibinfo
  {author} {\bibfnamefont {I.}~\bibnamefont {Souza}}, \bibinfo {author}
  {\bibfnamefont {D.}~\bibnamefont {Vanderbilt}}, \ and\ \bibinfo {author}
  {\bibfnamefont {N.}~\bibnamefont {Marzari}},\ }\href@noop {} {\bibfield
  {journal} {\bibinfo  {journal} {Computer physics communications}\ }\textbf
  {\bibinfo {volume} {178}},\ \bibinfo {pages} {685} (\bibinfo {year}
  {2008})}\BibitemShut {NoStop}%
\bibitem [{\citenamefont {Brooks}(1955)}]{brooks1955theory}%
  \BibitemOpen
  \bibfield  {author} {\bibinfo {author} {\bibfnamefont {H.}~\bibnamefont
  {Brooks}},\ }in\ \href@noop {} {\emph {\bibinfo {booktitle} {Advances in
  electronics and electron physics}}},\ Vol.~\bibinfo {volume} {7}\ (\bibinfo
  {publisher} {Elsevier},\ \bibinfo {year} {1955})\ pp.\ \bibinfo {pages}
  {85--182}\BibitemShut {NoStop}%
\bibitem [{\citenamefont {Chattopadhyay}\ and\ \citenamefont
  {Queisser}(1981)}]{chattopadhyay1981electron}%
  \BibitemOpen
  \bibfield  {author} {\bibinfo {author} {\bibfnamefont {D.}~\bibnamefont
  {Chattopadhyay}}\ and\ \bibinfo {author} {\bibfnamefont {H.}~\bibnamefont
  {Queisser}},\ }\href@noop {} {\bibfield  {journal} {\bibinfo  {journal}
  {Reviews of Modern Physics}\ }\textbf {\bibinfo {volume} {53}},\ \bibinfo
  {pages} {745} (\bibinfo {year} {1981})}\BibitemShut {NoStop}%
\bibitem [{\citenamefont {Bassani}\ and\ \citenamefont
  {Parravicini}(1975)}]{parravicini1975g}%
  \BibitemOpen
  \bibfield  {author} {\bibinfo {author} {\bibfnamefont {F.}~\bibnamefont
  {Bassani}}\ and\ \bibinfo {author} {\bibfnamefont {G.~P.}\ \bibnamefont
  {Parravicini}},\ }\href@noop {} {\emph {\bibinfo {title} {Electronic States
  and Optical Transitions in Solids}}}\ (\bibinfo  {publisher} {Pergamon
  press},\ \bibinfo {address} {New York},\ \bibinfo {year} {1975})\BibitemShut
  {NoStop}%
\end{thebibliography}%

\end{document}